\documentclass[a4paper,pra,superscriptaddress,amsmath,amssymb,mathrsfs]{revtex4}
\usepackage{graphicx}
\usepackage{braket}
\usepackage{hyperref}
\hypersetup{colorlinks=true,linkcolor=blue,citecolor=blue,urlcolor=blue}

\begin{document}

\title{Rational pulse design for enantiomer-selective microwave three-wave mixing}

\author{M. Leibscher}
\affiliation{Theoretische Physik, Universit\"at Kassel, Heinrich-Plett-Stra{\ss}e 40, 34132 Kassel, Germany}
\affiliation{Dahlem Center for Complex Quantum Systems and Fachbereich Physik, Freie Universit\"at Berlin, Arminallee 14, 14195 Berlin, Germany}
\author{J. Kalveram} 
 \affiliation{Theoretische Physik, Universit\"at Kassel, Heinrich-Plett-Stra{\ss}e 40, 34132 Kassel, Germany}
\author{C. P. Koch}
 \affiliation{Theoretische Physik, Universit\"at Kassel, Heinrich-Plett-Stra{\ss}e 40, 34132 Kassel, Germany}
 \affiliation{Dahlem Center for Complex Quantum Systems and Fachbereich Physik, Freie Universit\"at Berlin, Arminallee 14, 14195 Berlin, Germany}

\date{\today}
 
 \begin{abstract}
   Microwave three-wave mixing allows for enantiomer-selective excitation of randomly oriented chiral molecules into rotational states with different energy. The random orientation of molecules is reflected in the degeneracy of the rotational spectrum with respect to the orientational quantum number $M$ and reduces, if not accounted for,  enantiomer-selectivity. Here, we show how to design pulse sequences with maximal enantiomer-selectivity from an analysis of the $M$-dependence of the Rabi frequencies associated with rotational transitions induced by resonant microwave drives.
   We compare different excitations schemes for rotational transitions and show that maximal enantiomer-selectivity at a given rotational temperature is achieved for synchronized three-wave mixing with circularly polarized fields.
\end{abstract}

\maketitle

\section{Introduction}
Chiral molecules cannot be superimposed with their mirror image by rotation and translation; they exist in left- and right-handed forms called enantiomers. While the two enantiomers typically have different bio-chemical behavior, they share almost the same physical properties; in particular, they have practically identical spectra. Several techniques for the detection of enantiomers  exploiting the interaction of the molecules with electromagnetic radiation have recently been developed. {\color{black} Since these techniques do not rely on the inherently weak interaction with the magnetic field of the radiation, they have a sufficiently high sensitivity for applications in the gas phase.} Among these are ultrafast spectroscopies based on photoelectron circular dichroism~\cite{LuxAngewandte12,CireasaNatPhys15}, high-harmonic generation~\cite{BakyushevaPRX18,NeufeldPRX19}, enantiomer-selective control of molecular rotation~\cite{TutunnikovJPCL18,MilnerPRL2019,TutunnikovPRA2020} and
resonant phase-sensitive microwave three-wave mixing~\cite{PattersonNature13,ShubertAngewandte14,LobsigerJPCL15,Domingos2020}.
These techniques are based on light-matter interaction in the dipole approximation, where the enantiomer-selective observable arises as a triple product of molecule-specific vectors which changes sign under exchange of the two enantiomers, independent of the molecular orientation~\cite{BychkovJETP01,OrdonezPRA18}. In addition to detecting enantiomeric excess,
microwave three-wave mixing (3WM) can also be used to selectively excite enantiomers to different energy levels~\cite{EibenbergerPRL17,PerezAngewandte17,PerezJPCL18,Lee21}.
This can serve as a precursor for the separation of enantiomers and the preparation of an enantio-pure sample out of a racematic mixture \cite{KralPRL01, FrishmanJCP03}. 
The enantiomer-selective excitation proceeds in a cyclic way involving three rotational states, corresponding to the enantiomer-selective triple product of the three non-vanishing Cartesian components of the molecular dipole moment, $\mu_a$, $\mu_b$ and $\mu_c$, of which one changes sign under exchange of the two enantiomers~\cite{HirotaPJA12}. The sign change causes constructive interference for one enantiomer and destructive interference for the other~\cite{KralPRL01,KralPRL03,HirotaPJA12,LehmannJCP18,YePRA2018,Leibscher19,YongPRA2021}.

In order to distill a single enantiomer from a racemic mixture, an  enantiomer-selectivity of close to 100\% is required in the state transfer to the separate energy levels. 
In experiments, the efficiency is mainly limited by two factors.
One is the temperature of the sample, i.e., the rotational states which are addressed in a three-wave mixing process are typically thermally populated.  First demonstrations of enantiomer-selective state transfer chose one of the rotational levels with a large thermal weight as the starting point for the three-wave mixing~\cite{EibenbergerPRL17,PerezAngewandte17,PerezJPCL18}. Since thermal population in the excited states of the cycle cannot be coherently coupled, the contrast is reduced.
The second limitation is due to degeneracies within the rotational spectrum. Denoting the rotational quantum number by $J$, every energy level of a rigid asymmetric top consists of $2J+1$ states with different values of the orientational quantum number $M$. As a result, some of the parallel cycles fail to close, and cycles with different $M$ involve different Rabi frequencies.
This limits the efficiency of enantiomer-selective population transfer, even in the absence of thermal population in the excited states \cite{LehmannJCP18}.

The problem of temperature can be solved by addressing levels which are sufficiently excited such that their thermal population vanishes~\cite{Leibscher19,Zhang20}. Alternatively, the thermal population in the excited levels can be eliminated prior to the three-wave mixing, for example by optical pumping of an electronic transition \cite{Lee21}. 
The limitation due to the orientational degeneracy can also be mitigated---it requires a sufficiently large number of electric fields that break the corresponding symmetry~\cite{Leibscher20}. This has been shown by analyzing the controllability of finite dimensional subsystems of quantum asymmetric tops~\cite{Leibscher20,Pozzoli21}.  Such an analysis allows one to determine 
the minimal number of microwave fields which is necessary to control enantiomer selective rotational dynamics~\cite{Leibscher20}. It also establishes the polarization directions and the frequencies of the required control fields. However, it cannot determine the actual pulse shapes, or the order of pulses in a sequence of microwave excitations which separates the enantiomers of a racemic mixture into different rotational states.

In the present study, we show how to derive pulse sequences with maximal enantiomer-selectivity from a quantitative analysis of the population dynamics of degenerate rotational states induced by resonant microwave fields. Forfeiting complete controllability, we identify simpler pulse sequences than those of Ref.~\cite{Leibscher20} which nevertheless lead to full enantiomer selectivity. We furthermore show how the analysis of the rotational dynamics allows us to extend the excitation schemes of Ref.~\cite{Leibscher20} and predict the pulse parameters, in particular the duration of the required pulses, for different experimental conditions and different molecular species. 

The paper is organized as follows: In Section~\ref{sec:asymtop}, we summarize the properties of a rigid asymmetric top and its interaction with microwave fields. For a better understanding of the underlying rotational dynamics, we recall how rotational dynamics induced by a single microwave field depends on the quantum number $M$ of the rotational states (Section~\ref{sec:Mdependence}). These insights can be utilized to construct sequences of microwave pulses that result in complete enantio-selective populations transfer despite the presence of degenerate states, independent of the general controllability of the system. In Section~\ref{sec:completeselection}, we present two examples: in Subsection~\ref{subsec:linearfields}, we show that a combination of three different three-wave mixing cycles addresses all degenerate initial states and leads to complete enantio-selection. In Subsection~\ref{subsec:circularfileds} we explore enantio-selective population transfer with circular polarized microwave fields. The effects of rotational temperature and pulse duration are discussed in Subsection~\ref{subsec:temperature_accuracy}. In Section~\ref{sec:conclusions}, we summarize our results and conclude.

\section{Asymmetric top and its interaction with microwave radiation}
\label{sec:asymtop}
In general, rigid chiral molecules, {\color{black} i.e. chiral molecules in their electronic and vibrational ground state} are asymmetric top molecules with the rotational  Hamiltonian
\begin{equation}
  {\hat H}_{rot} =  A {\hat J_a}^2 + B {\hat J_b}^2 + C {\hat J_c}^2,
\end{equation}
where $\hat J_a$, $\hat J_b$ and $\hat J_c$ are the angular momentum operators with respect to the principle molecular axes, and $A > B > C$ are the rotational constants. The eigenfunctions of an asymmetric top are determined by
\begin{equation}\label{eq:Hrot}
H_{rot} |J,\tau,M \rangle = E_{J,\tau} |J, \tau, M \rangle\,.
\end{equation}
Since Eq.~\ref{eq:Hrot} does not have an analytical solution, the asymmetric top eigenfunctions are typically expressed in terms of symmetric top eigenfunctions which admit a closed form via Wigner D-matrices \cite{Zare88}.
The molecule becomes a prolate or oblate symmetric top with the
eigenfunctions $|J, K_a, M \rangle$ or $|J, K_c, M \rangle$ for $B=C$, respectively $A=B$. The symmetric top wavefunctions are characterized by the rotational quantum number $J=0, 1,2 ...$ and the quantum numbers $M=-J, -J+1,...,J$ and $K=-J, -J+1,...,J$ which describe the orientation with respect to a space-fixed and molecule-fixed axis, respectively.
The eigenfunctions of the asymmetric top are given by superpositions of symmetric top eigenstates,
\begin{equation}
   | J, \tau, M \rangle = \sum_K c_K^{J} (\tau) |J,K,M \rangle,  
   \label{asym_top}
\end{equation}
where $K$-states with the same $J$ and $M$ are mixed, and $\tau=-J,-J+1,...,J$. In rotational spectroscopy, the asymmetric top states are usually denoted by $|J_{|K_a|,|K_c|,M} \rangle$. We therefore use this notation to characterize the asymmetric top eigenstates. The two notations relate to each other as follows: For a given $J$, the asymmetric top states with lowest energy can be denoted either by $|J_{|K_a|=0,|K_c|=J,M} \rangle$ or, using the quantum number $\tau$, by  $|J,\tau=-J,M \rangle$, the ones with largest energy by $|J_{|K_a|=J,|K_c|=0,M} \rangle$ or $|J,\tau=J,M \rangle$. The states in between can be matched accordingly.
The eigenenergies $E_{J,\tau}$ of an asymmetric top do not depend on the quantum number $M$ and thus each rotational level is $2J+1$-fold degenerate.

The interaction of the molecules with an electromagnetic field in the electric dipole approximation is given by
\begin{equation}
   {\hat H}_{int} = - {\hat{\vec \mu}} \cdot {\vec E(t)}
 \end{equation}
 with the electric field 
 \begin{equation}
   {\vec E(t)} = {\vec e} E(t) \cos (\omega t + \phi)\,,
 \end{equation}
 where ${\vec e}$ denotes the polarization direction, $E(t)$ the temporal shape, $\omega$ the frequency and $\phi$ the phase of the field.
 Note that ${\hat{\vec \mu}}^T = ({\hat \mu}_x, {\hat \mu}_y, {\hat \mu}_z )$ is the molecular dipole moment in space-fixed coordinates. The transformation of the interaction Hamiltonian into molecule fixed coordinates is given by \cite{Zare88},

\begin{eqnarray}
 	\hat\mu_x &=& \frac{\mu_a} {\sqrt{2}}  \left( D_{-10}^1  - D_{10}^1 \right)  + \frac{\mu_b}{ 2} \left ( D_{11}^1 - D_{1-1}^1  - D_{-11}^1 + D_{-1-1}^1\right)  \nonumber \\ &-&  i \frac{\mu_c}{2} \left ( D_{11}^1 + D_{1-1}^1  - D_{-11}^1 - D_{-1-1}^1\right), \nonumber \\
 	\hat\mu_y  &=& -i \frac{\mu_a}{\sqrt{2}}  \left( D_{-10}^1  + D_{10}^1 \right)  + i \frac{\mu_b}{ 2} \left ( D_{11}^1 - D_{1-1}^1  + D_{-11}^1 - D_{-1-1}^1\right)  \nonumber \\ &+& \frac{\mu_c}{2} \left ( D_{11}^1 + D_{1-1}^1  + D_{-11}^1 + D_{-1-1}^1\right), \nonumber \\
 	\hat \mu_z  &=&  \mu_a D_{00}^1 - \frac{\mu_b}{\sqrt{2} } \left ( D_{01}^1 - D_{0-1}^1 \right) +  i \frac{\mu_c^{(\pm)}}{\sqrt{2}} \left ( D_{01}^1 + D_{0-1}^1 \right),
 	\label{mu_projection}
 	\end{eqnarray}  
where $D_{MK}^J$ denote the elements of the Wigner $D$-matrix. We consider the interaction of an asymmetric top with microwave radiation and assume that the frequency is resonant to a particular rotational transition, i.e. $\omega = E_{J',\tau'} -  E_{J,\tau}$. We assume that only rotational states with $E_{J',\tau'}$ and 
$E_{J,\tau}$ are addressed by the interaction. In broadband microwave spectroscopy, this assumption is typically justified for frequencies larger than about $50\,$MHz~\cite{PattersonNature13,ShubertAngewandte14,EibenbergerPRL17,PerezAngewandte17,PerezJPCL18}.

In order to investigate the population transfer between the rotational states induced by microwave pulses, we numerically solve the time-dependent Schr\"odinger equation, 
\begin{equation}
i \hbar \frac{\text{d}}{\text{d} t} |\psi(t) \rangle = \left( H_{rot} + H_{int}(t) \right) |\psi(t) \rangle\,,
\label{TDSE}
\end{equation}
using the Chebychev propagation technique~\cite{Kosloff_Review1994} and 
the basis of the asymmetric top eigenfunctions, i.e.,
\begin{equation}
 |\psi (t) \rangle = \sum_{J,\tau,M} a_{J,\tau}^M(t) |J, \tau, M \rangle.
\end{equation}
 The transition matrix elements between two asymmetric top states can be expressed in terms of those of the symmetric top eigenstates,
\begin{eqnarray}
  \langle J'' \tau'' M'' | D_{MK}^1 | J' \tau' M' \rangle = \sum_{K',K''} 
    c_{K'}^{J'} \left ( c_{K''}^{J''} \right )^\ast \langle J'' K'' M'' |D_{MK}^1|J' K' M' \rangle
\label{transition_asym}
 \end{eqnarray}
 with    
\begin{eqnarray}
   \langle J'', K'', M'' | D^1_{MK} | J', K', M' \rangle &=& \sqrt{2 J'' +1} \sqrt{2 J' +1} (-1)^{M''+K''}  \nonumber \\
  &\times&  \left(    \begin{array}{ccc}  J' & 1& J'' \\ M' & M & -M'' \end{array} \right)
    \left( \begin{array}{ccc} J' & 1 & J'' \\ K' & K & -K'' \end{array} \right)\,.
    \label{w3j}
\end{eqnarray}
The $M$-dependence of the transition matrix elements is due to the first  Wigner $3j$-symbol in Eq.~(\ref{w3j}).
While this dependence is well-known, it is useful to recall the corresponding rotational dynamics. This will allow us to better understand microwave three-wave mixing and to construct fully enantiomer-selective pulse sequences in the presence of degenerate states.
In Section~\ref{sec:Mdependence}, we therefore first discuss how the $M$-dependence of the transition matrix elements and the polarization of the microwave pulses affect the population transfer between the rotational states. In Section~\ref{sec:completeselection}, we then apply these results and design pulse sequences that allow for complete enantiomer-selective population transfer despite the degeneracy of the rotational spectrum.

\section{M-dependence of the population transfer between rotational states}
\label{sec:Mdependence}

\begin{figure}
  \centering
  \includegraphics[width=16cm]{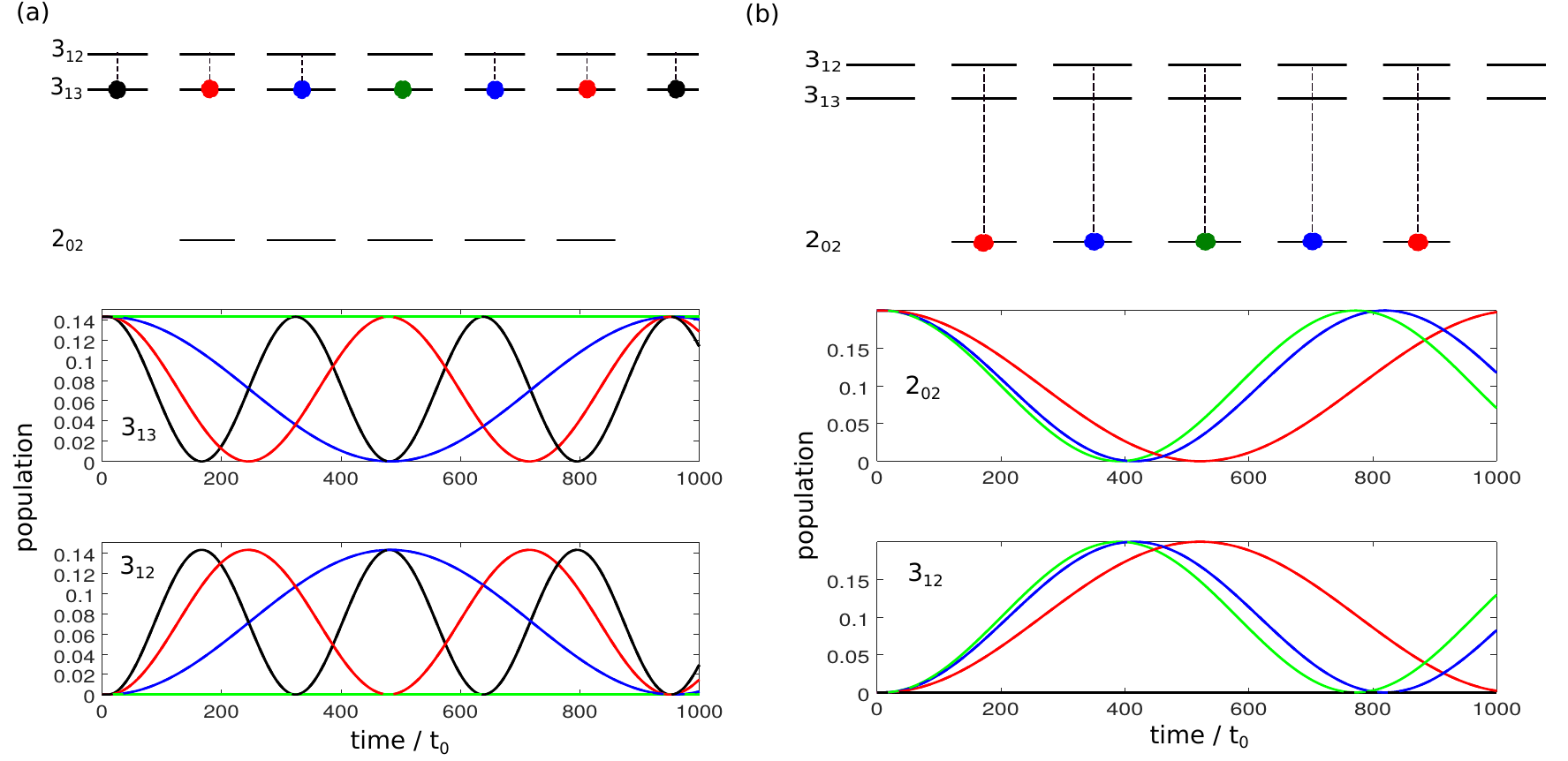}
  \caption{Population transfer between the rotational states $|3_{13M} \rangle$ and $|3_{12M} \rangle$ (a) and $|2_{02M} \rangle$ and $|3_{12M} \rangle$ (b) driven by a $z$-polarized microwave field.
    The rotational manifolds are depicted in the top panels. Colored circles represent the initial states. The lower panels show the population of the rotational states. Green, blue, red, and black lines correspond to states with $M=0$, $|M|=1$ and $|M|= 2$ and $|M|=3$ (only in panel (a))}, respectively.
  \label{transition_z}
\end{figure}

In the following, we consider a rotational subsystem consisting of the asymmetric top states $|2_{02M} \rangle$, $|3_{13M} \rangle$ and $|3_{12M} \rangle$. 
Transitions between these rotational states have been utilized in microwave experiments for enantiomer selective excitation \cite{PerezAngewandte17}. 
To recall how population transfer between rotational states depends on the orientational quantum number and for the sake of clarity, we assume in this subsection that a single microwave pulse interacts with the asymmetric top. 
Moreover, only one rotational level is taken to be occupied initially such that
 \begin{equation}
 \rho(t=0) = \frac{1}{2J_0+1} \sum_{M=-J_0}^{J_0} |J_0,\tau_0,M\rangle \langle J_0, \tau_0, M|\,.
  \label{rho_initial_single_pulse}
 \end{equation}
 Figure~\ref{transition_z} shows the rotational dynamics $|a_{J,\tau}^{M}|^2$ for the different initial states during the interaction with a linearly polarized field along the laboratory-fixed $z$-axis. 
We assume microwave fields with constant amplitude and with frequencies $\omega= E_{3_{12}} - E_{3_{13}}$ in Fig.~\ref{transition_z}(a) and $\omega= E_{3_{12}} - E_{2_{02}}$ in Fig.~\ref{transition_z}(b).
The relevant rotational states are sketched on the top, with the initially occupied states indicated by colored circles. Since for a $z$-polarized field, only transitions with $\Delta M = 0$ occur, the dynamics is divided into five individual two-level systems with Rabi frequencies 
 \begin{equation}
 \hbar \Omega = | \langle J', \tau', M | \mu_z E_0 | J, \tau, M \rangle |  \propto 
  \left | \left(    \begin{array}{ccc}  J & 1& J'\\ M & 0 & -M \end{array} \right) \right |.
   \label{Rabi_general}
 \end{equation}
 The Rabi frequencies, and thus the time required for complete population transfer depends on the quantum number $M$. For transitions with $\Delta J = 0$, cf. Fig.~\ref{transition_z}(a), 
 \begin{equation}
  \hbar \Omega \propto \left| \left(    \begin{array}{ccc}  J & 1& J\\ M & 0 & -M \end{array} \right) \right| = 
  \left | \frac{M}{\sqrt{(2J+1)(J+1)J}} \right|\,.
 \end{equation}
 The Rabi frequency is thus proportional to $|M|$, i.e., for states with $|M|=3$, three Rabi cycles occur at the time when one Rabi cycle is completed for states with $|M|=1$, as can be seen in Fig. \ref{transition_z}(a). Moreover, for $\Delta J = 0$, transitions with $M'=M''=0$ are forbidden. As a result, it is not possible to obtain complete population inversion between the levels $3_{13}$ and $3_{12}$ with a single resonant microwave pulse.
 For transitions with $\Delta J = 1$, cf. Fig.\ref{transition_z}(b), the Rabi frequencies are proportional to
 \begin{eqnarray}
\hbar \Omega \propto \left | \left(    \begin{array}{ccc}  J & 1& J+1\\ M & 0 & -M \end{array} \right) \right | = \left|
\frac{\sqrt{(J+M+1)(J-M+1)}}{\sqrt{(2J+3)(2J+1)(J+1)}} \right |.
\label{Rabi_z_deltaJ1}
\end{eqnarray}
Here, $\Omega$ is maximal and the Rabi cycle fastest for $M=0$. Rabi cycles with larger $M$ are slightly slower, as can be seen in Fig. \ref{transition_z}(b). The Rabi frequencies differ by irrational factors, namely $\Omega \propto \sqrt{9-M^2}$. Thus, only approximate population inversion for the two rotational levels can be obtained in finite time.

\begin{figure}
	\centering
	\includegraphics[width=16cm]{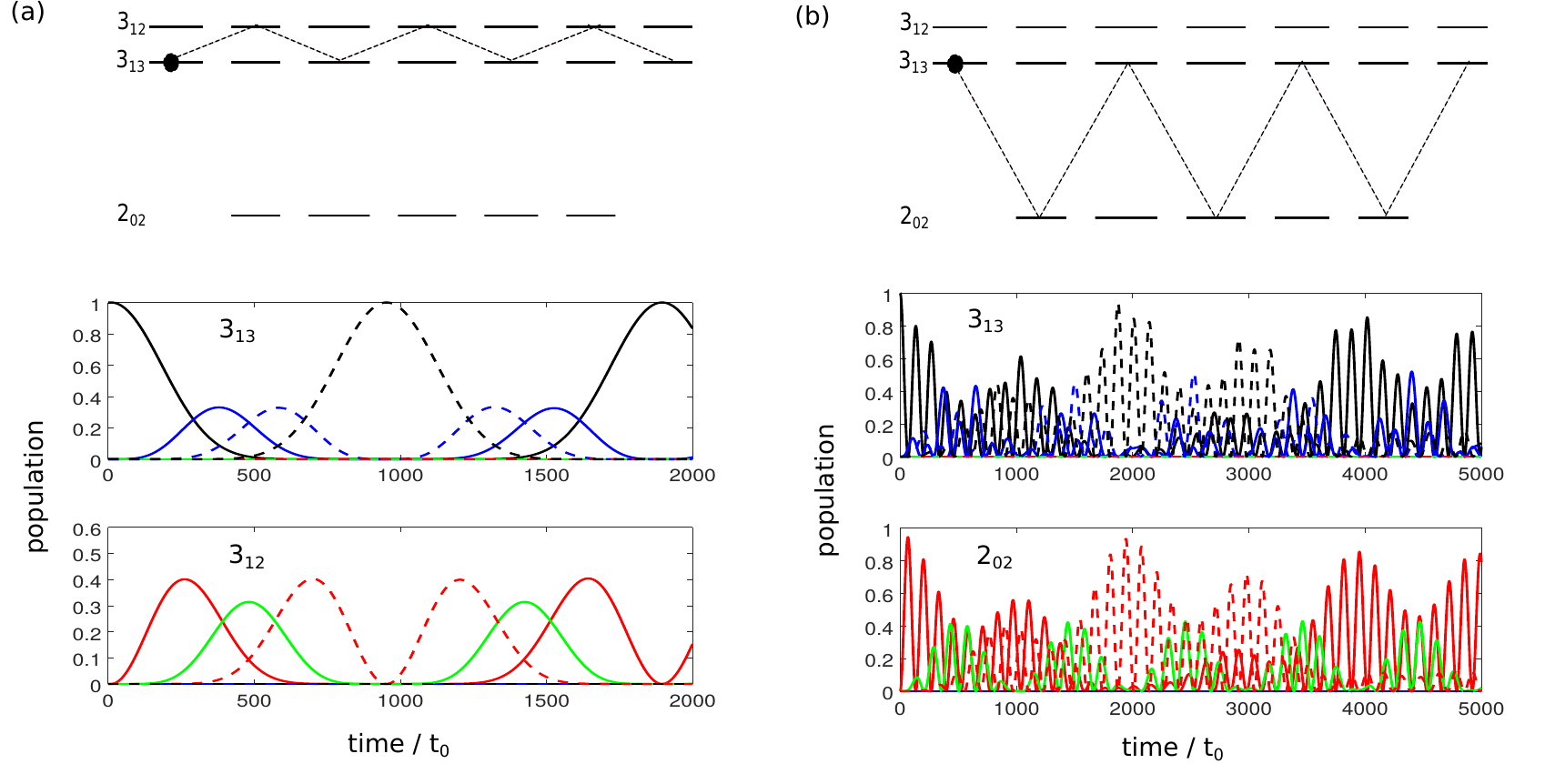}
	\caption{Same as Fig.~\ref{transition_z} but for an $x$-polarized field. Green lines correspond  to states with $M=0$, solid and dashed blue, red and black lines to states with $M=\mp 1$ and $M=\mp 2$ and $M=\mp 3$, respectively.}
	\label{transition_x}
\end{figure}

If we consider the interaction of an asymmetric top with a single microwave pulse, the polarization direction does not influence the rotational dynamics since we can always choose the quantization axis to be parallel to the polarization direction. However, three-wave mixing relies on the interaction of chiral molecules with three orthogonally polarized fields. To understand the underlying rotational dynamics, it is important to study the population transfer induced by fields which are polarized perpendicular to the quantization axis (here chosen to be the $z$-axis). We therefore consider the interaction of an asymmetric top with $x$-polarized microwave pulses in Fig.~\ref{transition_x}. 
For simplicity we consider only a single initial state, depicted by the black dots in Fig.~\ref{transition_x}(a) and (b). For the simulation shown in Fig. \ref{transition_x}(a), the initial state is $|3_{1,3,-3} \rangle$, and the frequency of the microwave pulse $\omega = E_{3_{12}} - E_{3_{13}}$. An $x$-polarized field induces transitions with $\Delta M = \pm 1$. Thus, the dynamics cannot be described by a two-level system. Instead, the field couples all states connected by the dashed lines in the top panel of Fig.~\ref{transition_x}(a) and population is transferred through the complete manifold of $M$-states. Nevertheless, complete population inversion occurs between the states $3_{1,3,-3}$ and $3_{1,3,3}$ with all states in between only partially populated.  This picture changes for transitions with $\Delta J = 1$, as shown in Fig.~\ref{transition_x}(b): Rapid oscillations between the states $3_{1,3,-3}$ and $2_{0,2,-2}$ with incomplete population transfer are observed before other states are substantially populated. In this case, population inversion between the states with maximal value of $|M|$
($3_{1,3,-3}$ and $3_{1,3,3}$) remains incomplete. The observation in 
both cases can be rationalized in terms of the three Rabi frequencies relevant to the seven states that are coupled by the $x$-polarized pulse (note the symmetry around $M=0$). In Fig.~\ref{transition_x}(a), these are $\Omega_1$, $\Omega_2 = 2 \Omega_1$ and $\Omega_3=3\Omega_1$, i.e.,  strictly periodic. In Fig.~\ref{transition_x}(b) the ratio between the three frequencies $\Omega_1$, $\Omega_2$ and $\Omega_3$ is irrational and thus not strictly periodic. Note that, for excitation with a single pulse, the population transfer induced by $x-$ and $y-$polarized fields is identical. The corresponding transition matrix elements only differ by sign \cite{Leibscher19}. This will become important for cyclic excitation with three orthogonal pulses.

\begin{figure}
	\centering
	\includegraphics[width=16cm]{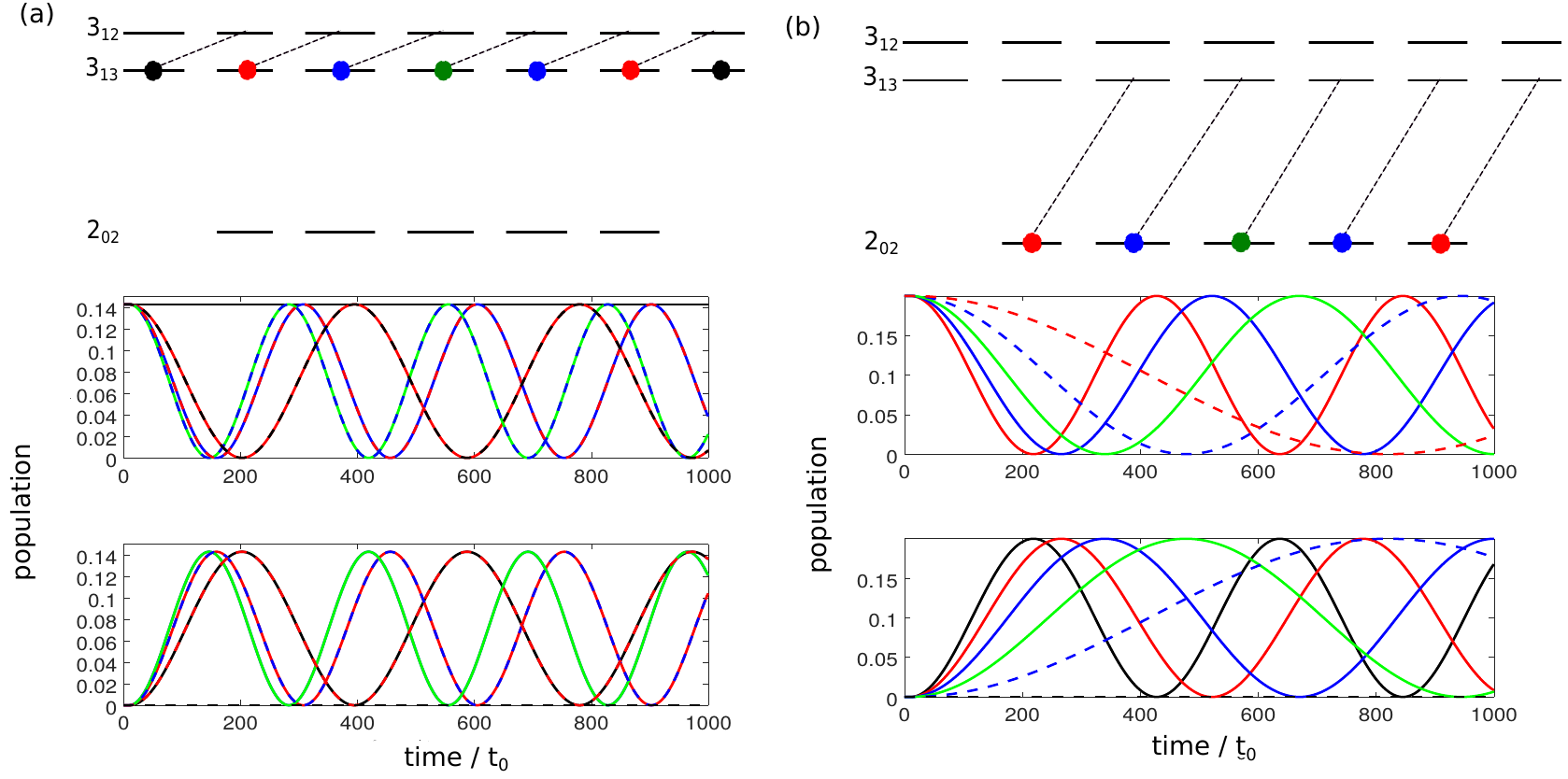}
	\caption{Same as Fig.~\ref{transition_z} but for a $\sigma_+$-polarized field. Green lines correspond  to states with $M=0$, solid and dashed blue, red and black lines to states with $M=\mp 1$ and $M=\mp 2$ and $M=\mp 3$, respectively.}
	\label{transition_sigma}
\end{figure}

Due to the spread of population over the complete $M$-manifold, excitation with a combination of $x$- $y$ and $z$-polarized fields does not result in closed three-level cycles and thus poses a challenge to three-wave mixing since
One way to overcome this problem is to use circularly polarized fields with polarization ${\vec e}_{\sigma_\pm}= \vec{e}_x \pm i {\vec e}_y$ instead of linearly polarized fields. The rotational dynamics resulting from interaction with  ${\sigma}_+$-polarized microwave pulses is shown in Fig. \ref{transition_sigma}. 
A field with ${\sigma}_+$-polarization allows for transitions with $\Delta M = + 1$ from the lower to the higher level and with $\Delta M = - 1$ for the reverse process. With such transitions, the rotational manifold decomposes into individual two-level systems. The $M$-dependence of the Rabi frequencies for right-circularly polarized radiation is given by 
\begin{eqnarray}
\hbar \Omega \propto \left | \left(    \begin{array}{ccc}  J & 1& J\\ M & 1 & -(M + 1) \end{array} \right) \right | = \left|
\frac{\sqrt{(J - M)(J + M+1)}}{\sqrt{(2J+2)(2J)J}} \right |.
\label{Rabi_sigma_deltaJ0}
\end{eqnarray}
for  transitions with $\Delta J = 0$ and by
\begin{eqnarray}
\hbar \Omega \propto \left | \left(    \begin{array}{ccc}  J & 1& J+1\\ M & 1 & -(M + 1) \end{array} \right) \right | = \left|
\frac{\sqrt{(J + M+2)(J + M+1)}}{\sqrt{(2J+3)(2J+2)(2J+1)}} \right |
\label{Rabi_sigma_deltaJ1}
\end{eqnarray}
for transitions with $\Delta J =1$. In both cases, for different values of $|M|$ the Rabi frequencies differ by irrational factors. Complete population transfer between two rotational levels can thus be achieved only approximately. 

In summary, the calculations presented in this section illustrate 
three mechanisms by which $M$-degeneracy affects rotational dynamics --- forbidden transitions, $M$-dependence of transitions-matrix elements and thus Rabi frequencies, and occurrence of sequential transitions for population transfer in case of $x$- and $y$-polarized pulses. With this knowledge, we can already draw conclusions for the design of pulses leading to complete enantiomer selective population transfer. The sequential transitions which occur in case of $x$- and $y$-polarized pulses complicate the design of pulse sequences. While  complete enantiomer-selective population transfer is possible when using only linearly polarized fields, it comes at the expense of rather complicated pulse sequences~\cite{Leibscher20}. A more straightforward way to construct a pulse sequence for complete enantiomer selective excitation is presented in Section \ref{subsec:linearfields}. Moreover, Fig.~\ref{transition_sigma} tells us that the simplest way to extend three-wave mixing to a manifold of degenerate states is to apply a combination of circularly polarized fields with a $z$-polarized field since this leads to a set of parallel three-level cycles. Due to the $M$-dependence of the Rabi-frequencies, parallel cycles need to be synchronized to achieve complete enantiomer selective excitation \cite{Leibscher20}. In contrast to excitation schemes with only linearly polarized fields, three-wave mixing using a combination of circularly and linearly polarized fields can be adopted in a straightforward manner to different rotational subsystems. In Section \ref{subsec:circularfileds}, we apply it to those rotational transitions in the carvone molecule which have been utilized in earlier microwave three-wave-mixing experiments \cite{PerezAngewandte17}.

\section{Complete enantiomer-selective excitation of degenerate rotational states}
\label{sec:completeselection}
In the following, we make use of the insights from Section~\ref{sec:Mdependence} to design pulse sequences which induce complete enenatiomer-selective excitation despite the degeneracy of rotational states. {\color{black} All results presented in this section are obtained by numerically solving the time-dependent Schr\"odinger equation (\ref{TDSE})}.
In Subsection~\ref{subsec:linearfields}, we design sequences of linearly polarized pulses to achieve complete enantiomer selectivity. As mentioned in
Section~\ref{sec:Mdependence}, the construction of closed cycles for complete enantiomer selection is particularly difficult in the presence of degenerate initial states if only linearly polarized pulses are used. We therefore consider the most simple rotational subsystem, i.e. the  manifold of $J=0$ and $J=1$, in this subsection.
Rotational subsystems with larger $J$ will be considered in
in Subsection \ref{subsec:circularfileds}, where synchronized circularly polarized pulses are applied. 
The effects of initial rotational temperature and of the pulse duration are discussed in Subsection~\ref{subsec:temperature_accuracy}.
 
\subsection{Pulse design using linearly polarized fields: Combination of three-wave mixing cycles}
\label{subsec:linearfields}
\begin{figure}
	\centering
	\includegraphics[width=\linewidth]{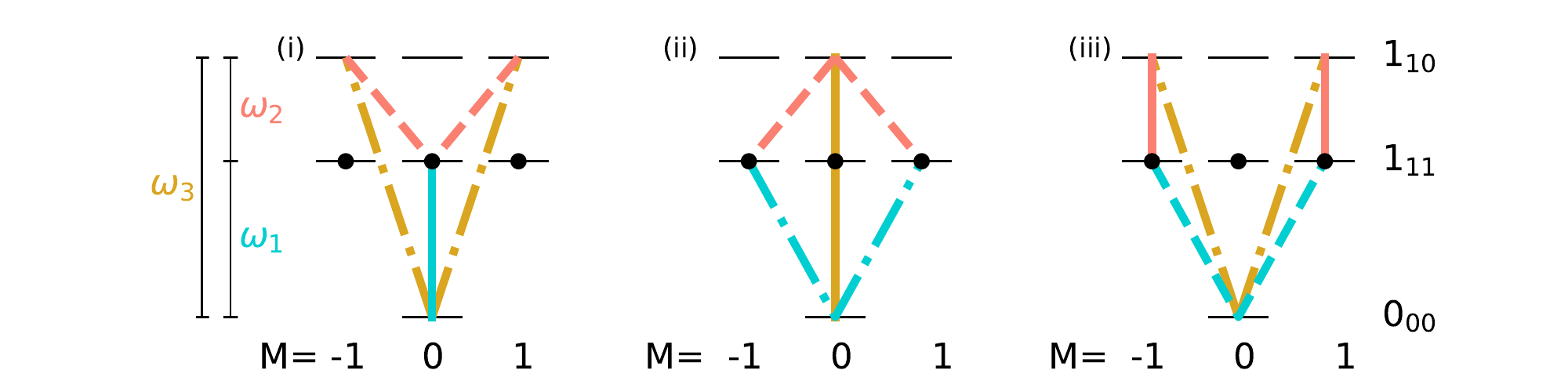}
	\caption{Three different three-wave-mixing schemes {\color{black} (i), (ii), and (iii)} for the rotational levels $0_{00}$, $1_{11}$ and $1_{10}$ with $x$-, $y$- and $z$-polarized pulses  indicated by dashed, dotted, and solid lines, respectively, the frequencies $\omega_1$, $\omega_2$ and $\omega_3$ by turquoise, orange and yellow lines, and the initial states by black circles.}
	\label{fig_schemaj1j2}
\end{figure}

The smallest subsystem with three rotational levels consists of the state with $J=0$, i.e., the non-degenerate rotational ground state denoted by $0_{00}$, and two excited rotational levels with $J=1$, namely $1_{11}$ and $1_{10}$, as shown in Fig. \ref{fig_schemaj1j2}. 
In a typical three-wave mixing pulse sequence, the first pulse creates a 50-50-coherence between the ground state and one of the  excited states. The second (twist) pulse transfers the complete excited state population to the second excited state. Finally, the third pulse induces the separation by causing constructive interference for one enantiomer and destructive interference for the other in the respective rotational state. Starting with the rotational ground state, interaction with such a sequence of three pulses, polarized in $x-$, $y-$, and $z-$ direction and with frequencies $\omega_1$, $\omega_2$ and $\omega_3$ induces complete enantio-selective population transfer \cite{Leibscher19}.
However, if the initial condition is given by
\begin{equation}
\rho(t=0) = \frac{1}{3} \sum_M |1_{11M} \rangle \langle 1_{11M}|,
\end{equation}
i.e., if the three degenerate states of level $1_{11}$ are populated, such a three-wave mixing pulse sequence cannot induce complete enantiomer selective excitation. This can be rationalized with the help of Fig.~\ref{fig_schemaj1j2}. A three-wave mixing cycle can be realized by three different combinations of $x$-, $y$-  and $z$-polarized microwave pulses, as seen in panels (i) - (iii). Here, the transitions induced by $x$-, $y$-, and $z$-polarized fields are marked by dashed, dotted, and solid lines, respectively, with the colors indicating the frequencies of the corresponding fields. The degenerate initial states are indicated by black circles. The three-wave mixing cycle (i) only affects the initial state with $M=0$, while the cycles (ii) and (iii) transfer the population from the initial states with $M=\pm1$.
Thus, only one or two of the initially populated states are part of each closed three-level cycle and therefore only part of the initial population can be selectively transferred to different rotational states.

\begin{figure}
  \centering
  \includegraphics{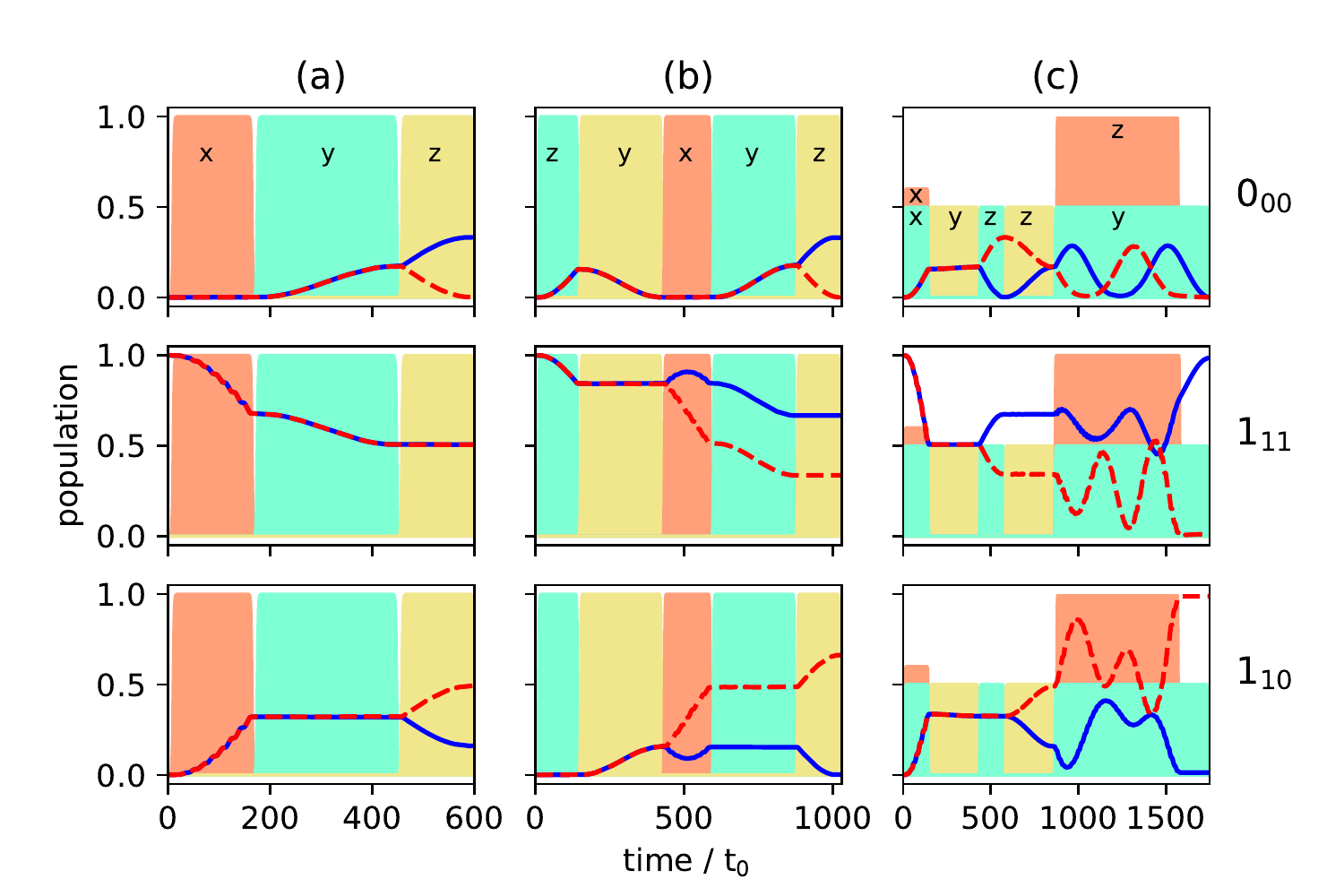}
  \caption{Population of the rotational levels $0_{00}$ (top), $1_{11}$ (middle) and $1_{10}$ (bottom) averaged over all $M$-states. The enantiomers $(+)$ and $(-)$ are presented by solid blue and dashed red lines. The envelopes of the microwave pulses are indicated by the turquoise ($\omega=\omega_1$), orange ($\omega=\omega_2)$) and yellow ($\omega=\omega_3$) shapes, and the polarization of the fields by the indices $x,y$ and $z$. (a): Excitation by the three microwave pulses indicated in Fig. \ref{fig_schemaj1j2} (i); (b): Excitation by a combination of the two three-wave mixing cycles depicted in Fig. \ref{fig_schemaj1j2} (i) and (ii); (c): Excitation by a combination of all three cycles (i), (ii) and (iii).}
  \label{3wm_J01}
\end{figure}
\begin{figure}
  \centering
  \includegraphics{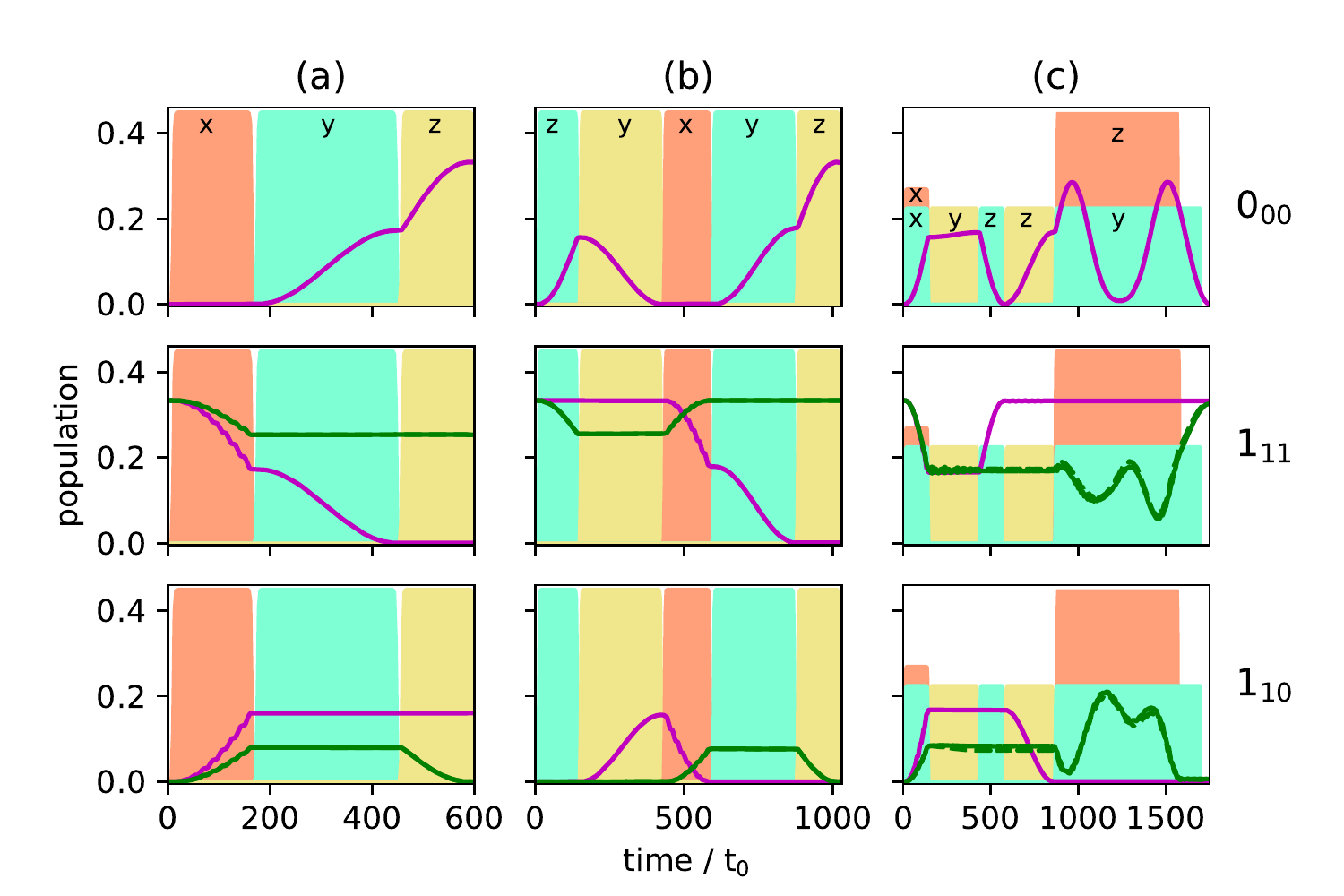}
  \caption{Population of the individual $M$-states of the rotational levels $0_{00}$ (top), $1_{11}$ (middle) and $1_{10}$ (bottom), shown for enantiomer $(+)$. The purple lines represent states with $M=0$, the green lines present $|M|=1$. The excitation schemes {\color{black} for panels (a), (b) and (c) are}  the same as in Fig. \ref{3wm_J01}.}
  \label{3wm_J01m}
\end{figure}
Figures~\ref{3wm_J01}(a) and \ref{3wm_J01m}(a) present the population transfer for the three-wave mixing cycle depicted in Fig.~\ref{fig_schemaj1j2}(i) with Fig. \ref{3wm_J01}(a) showing the time-dependent population of each rotational level averaged over the corresponding $M$-states for the two enantiomers (depicted by solid blue and dashed red lines), whereas the population of the individual $M$-states is shown in Fig. \ref{3wm_J01m}(a) for one of the enantiomers.
Cycle (i) results in complete enantiomer-selective population transfer for the initial state $|1_{110} \rangle$, whereas the initial states $|1_{11\pm 1} \rangle$ are  affected only by the first pulse, depicted in dashed orange lines in Fig. \ref{fig_schemaj1j2}(i). They are not part of a complete three-wave mixing cycle and thus no enantiomer-separation occurs for these initial states. As a result, the ground state $|0_{000} \rangle$ is populated only by a single enantiomer at final time, as shown in the top panel of Fig. \ref{3wm_J01} (a). In the first excited state, no enantiomer-separation is observed at all (middle panel), while only partial enantio-selectivity occurs in the second excited state, as shown in the bottom panel.

We quantify the overall selectivity by
\begin{equation}
S=\sum_{i=1}^{3} \Delta p_i(t_{final})\,,
  \label{selectivity}
\end{equation}
where $\Delta p_i(t) = |p_i^{(+)}-p_i^{(-)} | / |p_i^{(+)}+p_i^{(-)} |$ is the normalized population difference between the two enantiomers $(+)$ and $(-)$ in the three energy levels $i=0_{00}, 1_{11}, 1_{10}$.
For excitation with the three-wave mixing cycle (i), $S=0.33$. The same amount of selectivity can be obtained by cycles (ii) and (iii).

A better selectivity can be expected if cycle (i) is combined with cycles (ii) or (iii) since all initial states are then addressed by a closed cycle. As an example we combine cycles (i) and (ii). The resulting population transfer is depicted in Figs. \ref{3wm_J01}(b) and \ref{3wm_J01m}(b) where the pulse sequence starts with cycle (ii). The first pulse (dashed turquoise lines in Fig. \ref{fig_schemaj1j2}(ii)) affects only the initial states $|1_{11\pm 1} \rangle$ and creates a coherence between the states $|1_{11\pm 1} \rangle$ and $|0_{000} \rangle$. The second pulse (solid yellow line) transfers the ground state population to the second excited state. The third pulse, depicted by dashed orange lines in Fig. \ref{fig_schemaj1j2} (ii), closes the cycle and, at the same time, acts as the first pulse of cycle (i). By combination of the two three-wave mixing cycles, complete separation of the enantiomers is achieved in levels $0_{00}$ and $1_{10}$, while the selectivity in level $1_{11}$ is still incomplete. Overall, the selectivity increases to $ S= 0.66$. Although all initial states are part of closed cycles, complete enantiomer-selectivity cannot be obtained. This is due to the population transfer being induced by $x$- or $y$-polarized fields. As shown in Fig. \ref{transition_x}, an $x$-polarized field does not induce a Rabi oscillation between two levels, but rather spreads the population over all $M$-states. Thus, part of the population of the initial state $|1_{11-1} \rangle$ leaks from the cycle connecting $|1_{11-1} \rangle$, $|0_{000} \rangle$ and $|1_{100} \rangle$. The same holds for the population from the initial state 
$|1_{111} \rangle$.

Enantiomer-selective population transfer can be completed by adding the third excitation scheme from Fig. \ref{fig_schemaj1j2}. The resulting rotational dynamics is shown in Fig. \ref{3wm_J01}(c) and Fig.~\ref{3wm_J01m}(c). 
The corresponding pulse sequence consists of seven pulses: three pulses with $x$- $y$-, and $z$-polarization are resonant with the transition
$0_{00} \leftrightarrow 1_{11}$, two pulses, $x$- and $z$-polarized, are resonant with  $1_{11} \leftrightarrow 1_{10}$ and two pulses, $y$- and $z$-polarized are resonant with $0_{00} \leftrightarrow 1_{10}$. 
In order to obtain complete selectivity, it is important to properly synchronize the three cycles by using overlapping pulses and adjusting the field strengths.  At $t=0$, both enantiomers are assumed to populate only level $1_{11}$. At the end of the pulse sequence, enantiomer $(+)$ (blue lines) populates level $1_{11}$, while enantiomer $(-)$ (red lines) exists only in rotational states belonging to level $1_{10}$. The rotational  ground state is empty. By combining all three three-wave mixing cycles $S=0.98$ is obtained, i.e., almost 100 \% enantiomer-selectivity. Note that a systematic optimization of the pulse parameters will allow to push the enantiomer-selectivity  even closer to 100 \%.

The sequence achieving essentially complete enantiomer-selectivity, cf. Figs.~\ref{3wm_J01}(c) and ~\ref{3wm_J01m}(c), consists of seven different microwave fields. It is constructed by combining all the different three-wave mixing schemes that exist for this rotational subsystem. This should be compared to Ref.~\cite{Leibscher20}, where enantiomer-selective excitation for the same rotational subsystem has been identified by means of a controllability study. In particular, five different fields, i.e., fields with different combinations of frequency and polarization, were found to be sufficient for the system to be enantiomer-selective controllable~\cite{Leibscher20}. While the minimal number of different fields is given by controllability analysis, the actual pulse shapes or sequence of pulses has to be determined by other means. In Ref.~\cite{Leibscher20}, a sequence of 12 individual pulses (using five different combinations of frequency and polarization) has been shown to yield  complete enantiomer-selectivity. 
Controllability analysis on the one hand, and pulse design resulting from knowledge of the rotational dynamics are thus two complementary approaches to achieve a complete enantiomer-selective excitation (or any other desired target) in the presence of degeneracies in the rotational spectrum. The pulse sequence constructed here, while simpler than the pulse sequence found in Ref.~\cite{Leibscher20}, is challenging for current microwave experiments due to the need to carefully adjust the field intensities of overlapping pulses. Moreover, there is no automatic way to  transfer these results to rotational subsystems with larger $J$. To overcome these limitations, we consider in the following a different excitation strategy, namely the use of circularly polarized fields.

\subsection{Pulse design with circularly polarized fields}
\label{subsec:circularfileds}
\begin{figure}
  \centering
  \includegraphics[width=0.6\linewidth]{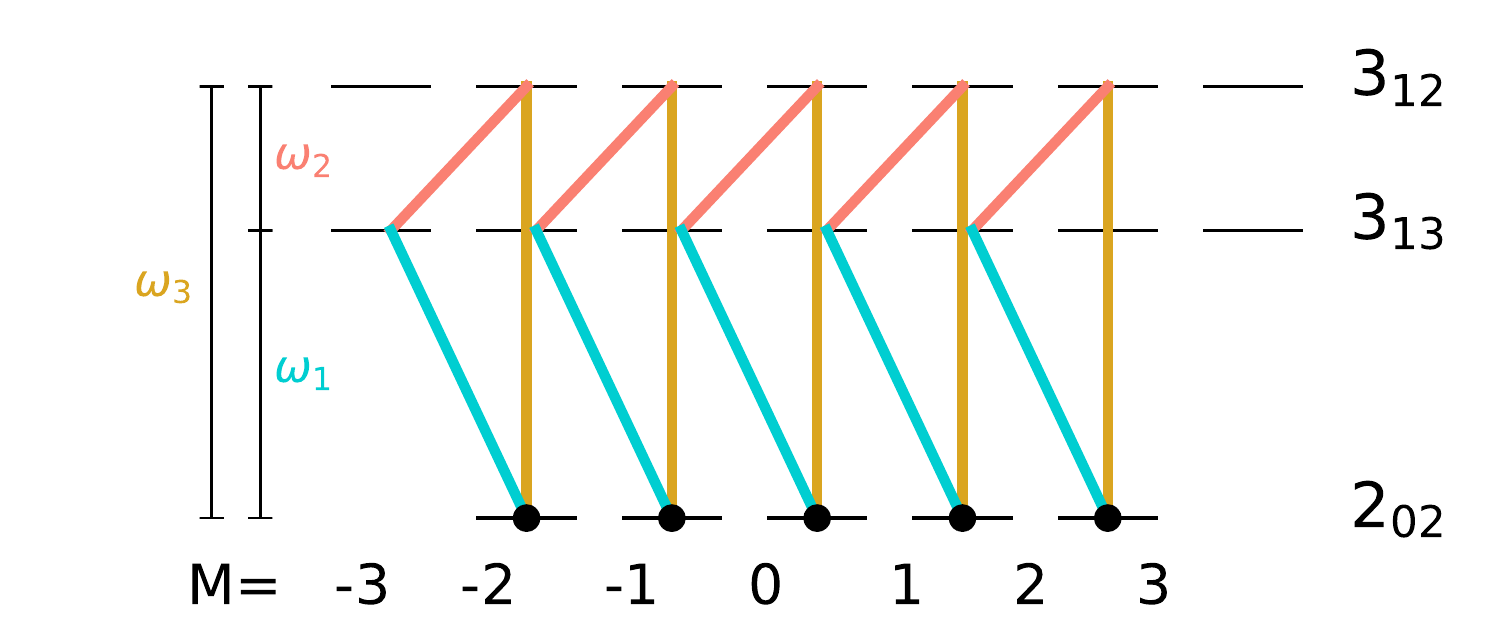}
  \caption{Three-wave mixing with circularly polarized fields for the rotational levels $2_{02}$, $3_{13}$ and $3_{12}$ with yellow, turquoise and orange lines indicating microwave fields with $\omega_3$ ($z$-polarization), $\omega_1$ ($\sigma_-$-polarization) and $\omega_2$ ($\sigma_+$-polarization). The initial states are indicated by black circles.}
  \label{scheme_J2J3}
\end{figure}

As discussed Section III, replacing the linear polarizations along $x$ and $y$ by circular polarizations prevents the spread of the initial population over the $M$-manifold. An excitation scheme using left- and right-circularly polarized pulses together with a $z$-polarized pulse has already been proposed in Ref.~\cite{Leibscher20}, where this combination of microwave fields was proven to lead to complete enantiomer-selectivity. Here, we extend this strategy to a set of rotational states for which microwave three-wave mixing was demonstrated experimentally~\cite{PerezAngewandte17}, namely the $2_{02}$, $3_{13}$ and $3_{12}$ states of carvone, depicted in Fig.~\ref{scheme_J2J3}.

\begin{figure}
  \center
  \includegraphics{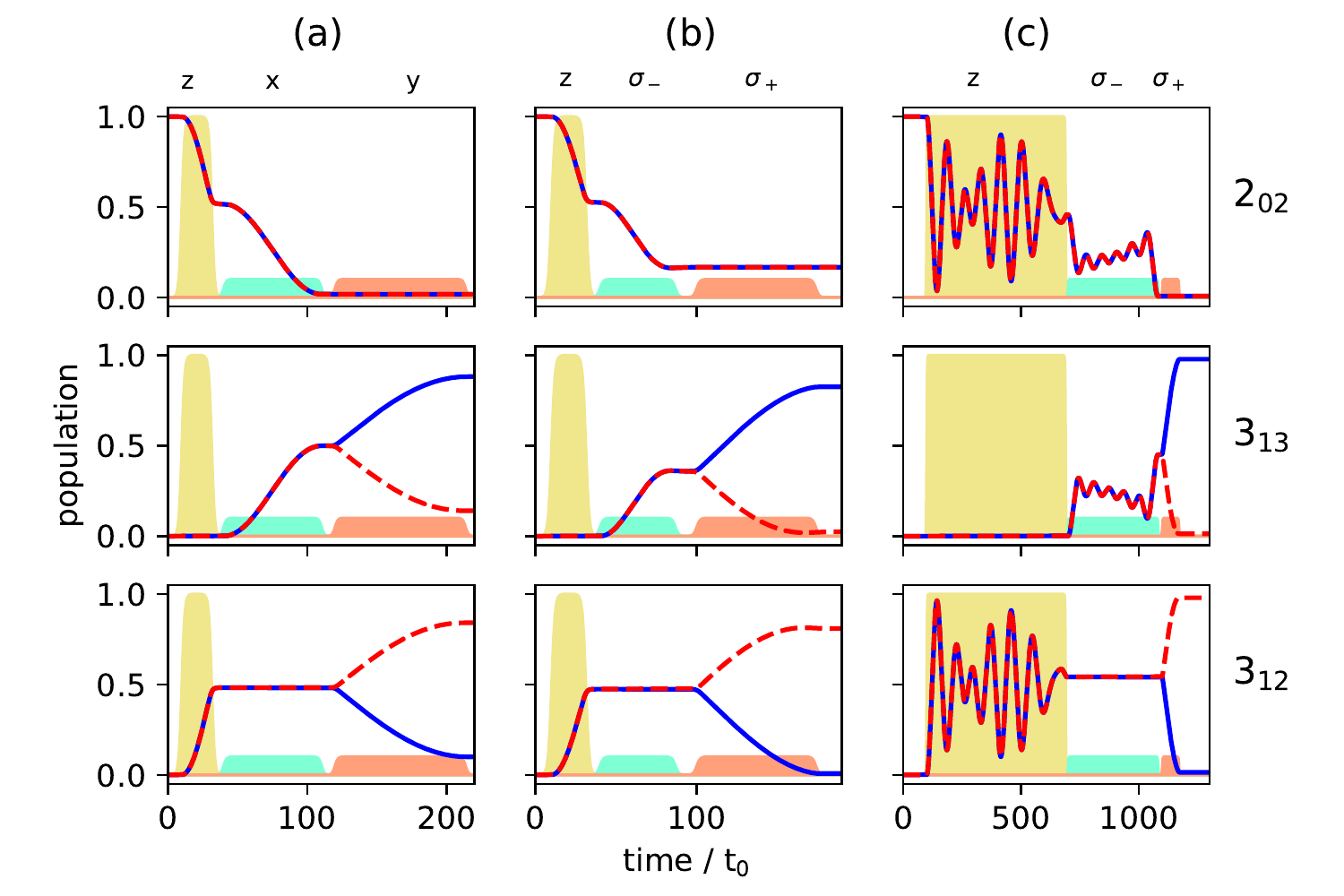}
  \caption{Population of the rotational levels $2_{02}$ (top), $3_{13}$ (middle) and $3_{12}$ (bottom) averaged over all $M$-states for excitation by a three-wave mixing scheme with $z$-, $x$-, and $y$-polarized fields (a),  with $z$-, $\sigma_-$- and $\sigma_+$-polarized fields (b) and  with $z$-, $\sigma_-$- and $\sigma_+$-polarized fields. Pulse durations are adjusted to allow for synchronization of the individual Rabi cycles. The two enantiomers are depicted by solid blue and dashed red lines, respectively. The envelopes of the microwave pulses are indicated by the turquoise ($\omega=\omega_1$), orange ($\omega=\omega_2$), and yellow ($\omega=\omega_3$) shapes. Note that the yellow pulse is 10 times as intense as the other pulses.}
	\label{3wm_J2J3}
\end{figure}
\begin{figure}
	\centering
	\includegraphics{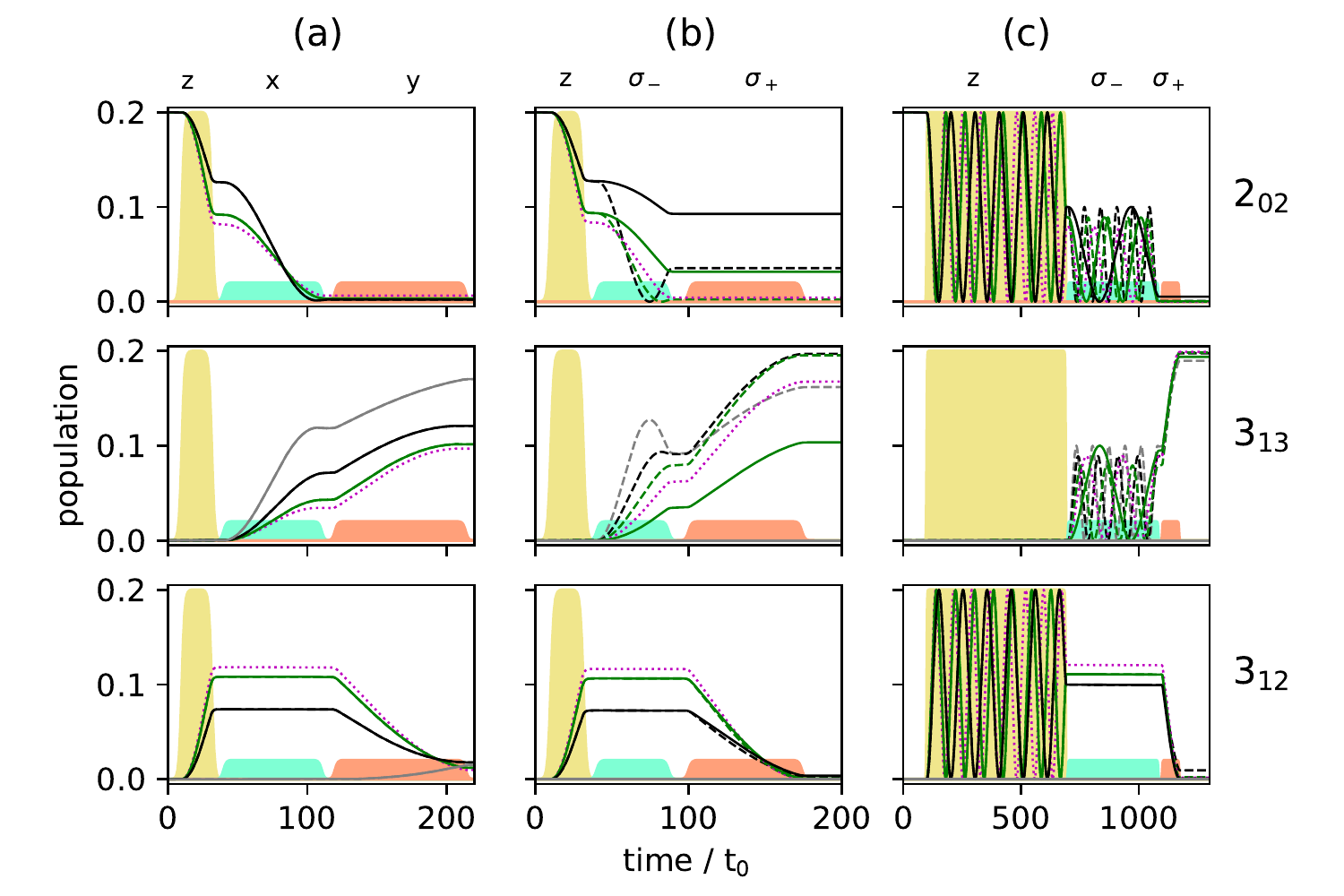}
	\caption{Population of the individual $M$-states for the rotational levels $2_{02}$ (top), $3_{13}$ (middle) and $3_{12}$ (bottom) for enantiomer $(+)$ for the same excitation schemes as in Fig. \ref{3wm_J2J3}. The solid and dashed gray, black and green lines present the states with $M=\mp 3$, $\mp 3$ and $\mp 1$, respectively. The states with $M=0$ are presented by the dotted purple lines.}
	\label{3wm_J2J3m}
\end{figure}
We assume that initially, all degenerate states of the lowest rotational level $2_{02}$ are equally populated, i.e.,
\begin{equation}
\rho(0) = \frac{1}{5} \sum_M |2_{02M} \rangle \langle 2_{20M} |\,,
\end{equation} 
c.f. the black circles in Fig. \ref{scheme_J2J3}, whereas levels $3_{13}$ and $3_{12}$ are empty.  Such an initial condition can be realized by choosing the two excited rotational states  in a vibrational state $\nu > 0$ \cite{Leibscher19,Zhang20} or by depleting excited rotational levels by laser excitation \cite{Lee21}.
Figures \ref{3wm_J2J3} and \ref{3wm_J2J3m} show the population transfer for different pulse sequences, with the average population of the three rotational levels for the two enantiomers plotted in Fig. \ref{3wm_J2J3}, while Fig. \ref{3wm_J2J3m} shows the population of the individual $M$-states for a single enantiomer. In panel (a) of Figs.~\ref{3wm_J2J3} and \ref{3wm_J2J3m}, we show, for reference, the population transfer for a three-wave mixing scheme with $x$-, $y$- and $z$-polarized fields which corresponds to an
enantiomer-selectivity of $S \approx 0.74$, with enantiomeric excess of enantiomer (+) and (-) in the levels $3_{13}$ and $3_{12}$, respectively and level $2_{02}$ empty. Using  $\sigma_+$ and $\sigma_-$-polarized  fields instead of the $x$- and $y$-polarized ones confines each of the initial states $|2_{02M}\rangle$ with $M=-2,...,2$ to a single 3-level cycle, as seen  in Fig.~\ref{scheme_J2J3}. We take the first pulse, as before, to be a $z$-polarized $\pi/2$-pulse, i.e., the pulse duration is chosen such that 50\% of the population (averaged over the $M$-states) is in level $3_{12}$ and 50\% remains in the ground state $2_{02}$. The second and third pulses are $\sigma_-$ and $\sigma_+$-polarized, respectively, with pulse durations determined as in a standard three-wave mixing scheme.
The resulting average population is shown in Fig.~\ref{3wm_J2J3}(b), and the population of the individual $M$-states can be seen in Fig.~\ref{3wm_J2J3m}(b).
The excitation scheme leads to almost complete enantiomer-selection for the levels $3_{12}$ and $3_{13}$. However, about 20\% of the population of both enantiomers remains in the lowest level $2_{02}$. This is already a clear advantage compared to the standard three-wave mixing scheme with only linearly polarized fields. One could, for example, obtain a purified sample by extracting the population of either level $3_{13}$ or $3_{12}$, each corresponding to a single enantiomer. The overall incomplete enantiomer-selectivity can be rationalized as follows: The first pulse leads to an average 50/50-coherence between the levels $2_{02}$ and $3_{13}$, c.f. Fig.~\ref{3wm_J2J3}(b). However, as also illustrated in Fig.~\ref{transition_z}, the Rabi frequencies of the two-level transitions with different $|M|$ are different, with transitions between $M=0$ states having the largest and transitions between $M=\pm 2$ states having the smallest Rabi frequency. This results in population transfer of more than 50\% for states with $M=0$ and $M=\pm 1$  and less than 50\% for $M=\pm 2$, c.f. Fig.~\ref{3wm_J2J3m}(b).

One can account for the different Rabi frequencies by adjusting the pulse duration such that every single $M$-state undergoes a $50/50$-population transfer. To achieve that, one has to wait for several Rabi cycles until all individual cycles are synchronized.  Similarly, the complete population transfer between the states $3_{12}$ and $3_{13}$ has to be synchronized. The pulse durations were determined by a simple parameter optimization using the NLopt software package~\cite{NLopt}. Pulse shapes and maximal field strengths were kept fixed. The resulting rotational dynamics is displayed in Figs. \ref{3wm_J2J3}(c) and \ref{3wm_J2J3m}(c). 
The fully synchronized three-wave mixing scheme with circularly polarized fields leads to almost complete enantiomer-selective excitation --- one enantiomer is entirely transferred to level $3_{12}$ (dashed red lines in Fig. \ref{3wm_J2J3}), while the second enantiomer ends up in level $3_{13}$ (blue lines in Fig. \ref{3wm_J2J3}). There is no population left in level $2_{02}$ and the overall enantiomer-selectivity amounts to $S=97\%$, 
limited -- according to Eqs.(\ref{Rabi_z_deltaJ1}), (\ref{Rabi_sigma_deltaJ0}) and (\ref{Rabi_sigma_deltaJ1}) -- by the Rabi frequencies for the different transitions differing by irrational factors.

While it is possible to synchronize the transitions such that a 50/50-coherence is obtained for all individual transitions with arbitrary accuracy, this comes at the expense of longer pulse durations. Eventually, the pulse durations are limited by the coherence time of the experiment, determined e.g. by collisions of the molecules in the sample with background gas. In practice, it is thus always necessary to find a compromise between accurate synchronization and pulse duration. We discuss the relation between accuracy and pulse duration in Subsection \ref{subsec:temperature_accuracy}. At the same time, we inspect another factor relevant in an experimental implementation. Most of the current microwave three-wave mixing experiments are carried out for rotational states with thermal population. In Section \ref{subsec:temperature_accuracy}, we thus discuss also the effects of the initial temperature.

\subsection{Rotational temperature and synchronization time}
\label{subsec:temperature_accuracy}
\begin{figure}
  \centering
  \includegraphics[width=0.8\linewidth]{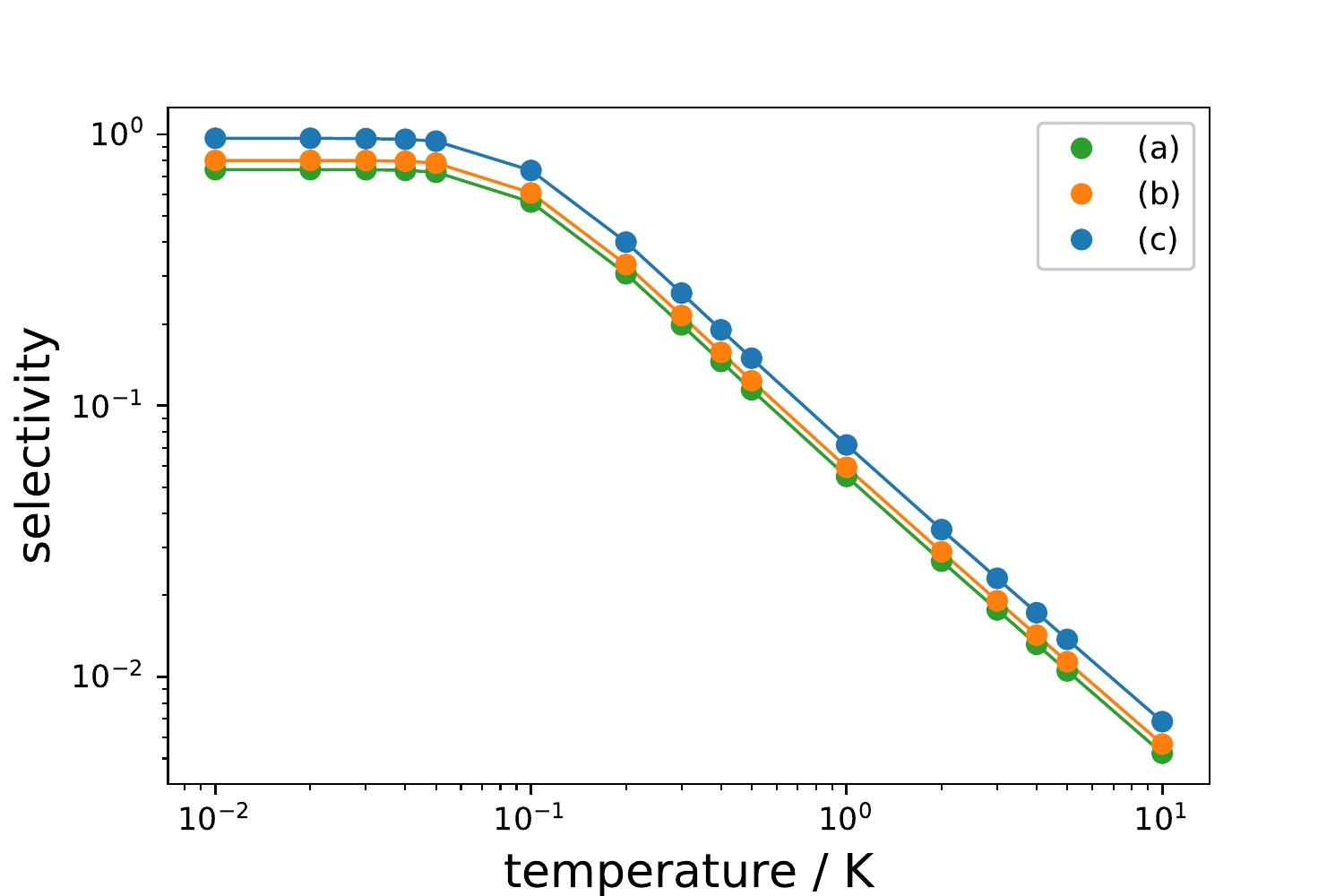}
  \caption{Selectivity $S$, Eq.(\ref{selectivity}) as a function of the initial rotational temperature. The green, blue and orange dots depict the selectivity for the excitation schemes shown in Fig. \ref{3wm_J2J3}(a), (b), and (c).}
  \label{temperature}
\end{figure}
So far, we have investigated the rotational dynamics of chiral asymmetric top molecules assuming that only the lowest rotational state of the relevant subsystem is initially populated. Typical microwave three-wave mixing experiments are carried out with thermal samples of chiral molecules~\cite{PerezAngewandte17,PerezJPCL18}. Then, the initial rotational temperature of the molecules has to be factored in. The initial density operator is given by
\begin{equation}
\rho(0) = \sum_{i=1}^3 \sum_{M=-J_i}^{J_i} p_{J_i,\tau_i}(T) |J_i,\tau_i,M \rangle \langle J_i, \tau_i, M |\,,
\end{equation} 
where the rotational levels are occupied according to the Boltzmann distribution,
\begin{equation}
  p_{J_i,\tau_i}(T) = \frac{1}{Q} \exp \left ( - \frac{E_{J_i,\tau_i}}{k_B, T}\right ) \,,
\end{equation}
with $Q$ defined by $\sum_{i=1}^{3} \sum_{M=-J_i}^{J_i}  p_{J_i,\tau_i}(T) = 1$ and $|J_1,\tau_1,M \rangle =|2,-2,M \rangle =|2_{02M} \rangle$,  $|J_2,\tau_2,M \rangle =|3,-2,M \rangle =|3_{13M} \rangle$ and  $|J_3,\tau_3,M \rangle =|3,-1,M \rangle =|3_{12M} \rangle$. 
In molecular beam experiments, rotational temperatures in the range of $1$ - $10$ K can be achieved, which result in non-neglibile initial thermal population of all rotational levels involved in the three-wave mixing cycle. Figure~\ref{temperature} shows the maximal selectivity for the  excitation schemes (a), (b), and (c) in Fig.~\ref{3wm_J2J3} in a semi-logarithmic plot for initial temperatures between $T=10\,$mK and $10\,$K.
For $ T < 0.1$ K, thermal occupation of the upper two levels becomes negligible and the selectivity approaches its maximal value, whereas it gets exponentially reduced for rotational temperatures $T > 0.1\,$K. Importantly,  the modified excitation scheme using synchronized circularly polarized pulses (Fig.~\ref{3wm_J2J3}(c)) results in a larger selectivity than standard three-wave mixing with linearly polarized fields (Fig.~\ref{3wm_J2J3}(a)) and three-wave mixing with circularly polarized fields without synchronization of the individual three-level systems (Fig.~\ref{3wm_J2J3}(b)) for every temperature. 

The maximal enantiomer-selectivity that can be achieved in practice with the optimal scheme employing synchronized circularly polarized fields also depends on how accurate the individual Rabi cycles are synchronized. In the following, we discuss the relation between accuracy and the pulse duration in more detail. Since we consider resonant excitation, the population of the lowest states is given by $p_{2_{02M}} = \cos^2 ( \Omega t )$, where the Rabi frequency $\Omega$, given in Eq.~(\ref{Rabi_general}), depends on $M$. Consider the first pulse in excitation scheme (c), which is a $z$-polarized field that drives a $\Delta J=1$ transition, i.e., the $M$-dependence of the Rabi frequency is given in Eq.(\ref{Rabi_z_deltaJ1}).  As shown in Fig.~\ref{transition_z}(b), the first pulse simultaneously drives $2J +1$ two-level transitions. In the bottom panel of Fig. \ref{3wm_J2J3m} (c), it can be seen that after the first pulse, the 50/50-coherence is reached for all M-stats within an accuracy of $10 \%$. Since the Rabi frequencies for different $M$-states differ by irrational numbers, a 50/50-coherence can be obtained with arbitrary accuracy by increasing the pulse duration.  
\begin{figure}
  \centering
  \includegraphics[width=0.6\linewidth]{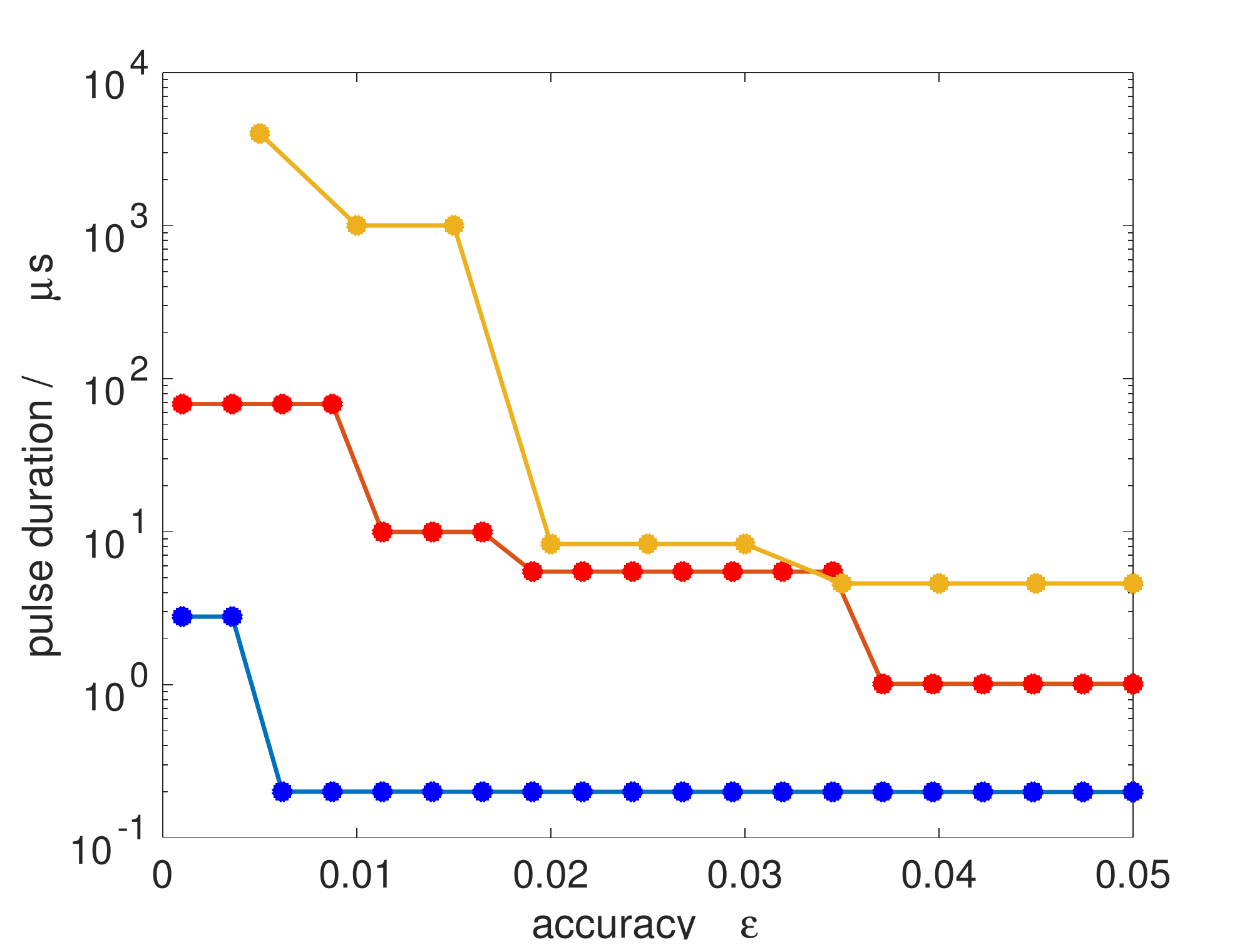}
  \caption{Pulse duration required for achieving a 50/50-coherence for all $M$-states within accuracy $\epsilon$ for transitions from $|1_{01M}\rangle$ to $|2_{11M}\rangle$ (blue dots), $|2_{02M} \rangle$ to $|3_{12M}\rangle$ (orange dots) and from  $|3_{03M}\rangle$ to $|4_{13M}\rangle$ (yellow dots) assuming a field intensity of $I=10$ W/cm$^2$.}
	\label{synchronization}
\end{figure}
In Fig. \ref{synchronization}, we show the pulse duration required to achieve a 50/50-coherence for all $M$-states within an accuracy $\epsilon$ defined by $p_{2_{02M}}= \left(\frac{1}{2} \pm \epsilon \right) p_{2_{02M}}(t=0)$. Transitions from $J=1$ to $J=2$ (blue dots), $J=2$ to $J=3$ (orange dots) and $J=3$ to $J=4$ (yellow dots) are considered, assuming a microwave field with an intensity of $10$ W/cm$^2$, comparable to field intensities used in current microwave experiments \cite{PerezAngewandte17}. A jump in the pulse duration occurs whenever the number of Rabi cycles has to be increased
to obtain a smaller value of $\epsilon$. Since the degeneracy increases with increasing $J$, the pulse duration required for very accurate synchronization increases. For a rotational subsystem with initial state $J=1$, a 50/50-coherence with an accuracy below one per cent can be achieved within less than  $1\,\mu$s. Starting with rotational states with $J=2$, one per cent accuracy already requires a pulse duration of about $10\,\mu$s. This is of the same order as typical coherence times in current microwave experiments.
Of course, the absolute value of the pulse duration can in principle be reduced by employing stronger pulses.

\section{Discussion and conclusions}
\label{sec:conclusions}

We have considered the problem that orientational degeneracy of randomly oriented chiral molecules impedes complete enantiomer selective excitation of asymmetric top molecules with resonant microwave three-wave mixing~\cite{LehmannJCP18}. Complementing earlier work using mathematical controllability analysis~\cite{Leibscher20}, we have shown how to solve this problem based on a detailed analysis of the rotational dynamics.
Our results are relevant to  microwave three-wave mixing experiments
employing rotational transitions between degenerate rotational states~\cite{EibenbergerPRL17,PerezAngewandte17,PerezJPCL18}. The excitation schemes derived here will improve the enantiomer selectivity in those experiments at the expense of replacing two of the linearly polarized by circularly polarized microwave fields with their duration tuned to synchronize transitions involving different $M$-dependent Rabi frequencies.
One option consists in combining all possible three-wave mixing cycles in a given rotational subsystem. This leads  to a simpler excitation scheme
compared to  Ref. \cite{Leibscher20}. However, since the construction does not guarantee controllability, it is not obvious whether it can be applied to arbitrary rotational subsystems. We have therefore also revisited the strategy derived from controllability analysis in Ref.~\cite{Leibscher20}, adapting it to those rotational states of carvone which have been addressed in   microwave three-wave mixing experiments \cite{PerezAngewandte17}.
To estimate the improvement in enantiomer selectivity that one can expect to observe in experiments, we have compared different excitations schemes at thermal conditions. In line with Ref.~\cite{Leibscher20}, we find that synchronized three-wave mixing with circularly polarized fields results in the best selectivity at a given temperature. Moreover, we have identified the pulse duration required for a desired accuracy of the rotational transfer. We emphasize that the excitation schemes discussed here are feasible with current microwave technology and  thus provide a promising route to increase the efficiency of enantiomer-separation of chiral molecules by microwave spectroscopy.

A recent three-wave mixing experiment \cite{Lee21} has followed a different path to circumvent loss of efficiency due to both thermal population in the excited rotational states and orientational degeneracy: The experiment addresses rotational levels with $J=0$ and $J=1$ with the excited rotational levels depleted prior to the three-wave mixing. Of course, in a typical molecular sample, only a small amount of molecules resides initially in the rotational ground state. In view of enantiomer separation with electric fields only, it will therefore be important to combine the optical depletion technique of Ref.~\cite{Lee21} with the pulse design presented here, in order to sequentially carry out several three-wave mixing cycles and thereby address a larger set of rotational states.

\begin{acknowledgments}
We would like to thank Karl Horn and Eugenio Pozzoli for helpful discussions. Financial support from the Deutsche Forschungsgemeinschaft through CRC 1319 ELCH is gratefully acknowledged.
\end{acknowledgments}

\bibliographystyle{apsrev4-1}

\end{document}